\documentclass[aps,pre,groupedaddress,amsmath,showpacs,twocolumn]{revtex4} 
\usepackage{graphicx,amssymb} 
\usepackage{bm} 
\usepackage{amsmath} 
\newcommand\beq{\begin{equation}} 
\newcommand\eeq{\end{equation}} 
\newcommand\beqa{\begin{eqnarray}} \newcommand\eeqa{\end{eqnarray}}   \newcommand{\nn}{\nonumber\\} \begin{document} 
\title{Inherent Rheology of a Granular Fluid in Uniform Shear Flow} \author{Andr\'es Santos} \email{andres@unex.es} \homepage{http://www.unex.es/fisteor/andres/} \author{Vicente Garz\'{o}} \email{vicenteg@unex.es} \homepage{http://www.unex.es/fisteor/vicente/} \affiliation{Departamento de F\'{\i}sica, Universidad de Extremadura, E--06071 Badajoz, Spain} \author{James Dufty} \email{dufty@phys.ufl.edu} \homepage{http://www.phys.ufl.edu/~dufty/} \affiliation{Department of Physics, University of Florida, Gainesville, FL 32611, USA}  
\date{\today}  \begin{abstract} { In contrast to normal fluids, a granular fluid under shear supports a steady state with uniform temperature and density since the collisional cooling can
compensate locally for viscous heating. It is shown that the hydrodynamic description of this steady state is inherently non-Newtonian.   As a consequence, the Newtonian shear viscosity cannot be determined from experiments or  simulation of uniform shear flow.  For a given  degree of inelasticity, the complete nonlinear dependence of the shear viscosity on the shear rate requires the analysis of the unsteady hydrodynamic behavior. The relationship to the Chapman--Enskog method to derive hydrodynamics is clarified using an approximate Grad's solution of the Boltzmann  kinetic equation}. \end{abstract}   
   
 \maketitle 
\section{Introduction\label{sec1}} Consider a fluid between parallel plates in the $x,z$ plane separated by a distance $L$  and with relative velocity $U$ along the positive $x$ direction. Given appropriate boundary conditions, the fluid undergoes simple shear with the local velocity field given by $u_{x}=ay,u_{y}=u_{z}=0$, where  $a=U/L$ is the constant shear rate. The work done by the plates on the fluid near the boundaries tends to increase the temperature locally due to its viscosity. For normal fluids, this is compensated by a heat flux toward the center with a corresponding temperature gradient characterizing Couette flow in the steady state. A quite different steady state is possible for granular fluids where both the temperature and density fields are spatially uniform, called {simple or} uniform shear flow (USF) \cite{C90,G03}. This is possible because the particle collisions in a granular fluid are inelastic and there is a continual loss of energy. This collisional cooling can compensate locally for the viscous heating so that no heat flux is generated. If the temperatures at the walls are not controlled the fluid autonomously seeks the temperature at which this exact balance between collisional cooling and viscous heating occurs. Otherwise, for fixed wall temperature one or the other will dominate leading again to Couette flow but with the curvature of the temperature field controlled by both mechanisms. This more general case has been discussed in detail elsewhere \cite{Tij} including USF as a special case. Here, attention will be limited to the special features of USF.   

{The steady USF has been extensively studied. Molecular dynamics and Monte Carlo simulations  \cite{WB86,C89,HS92,GT96,F00,AH01,AL03} have been performed to measure the dependence of the stress tensor on the  density and inelasticity. On the theoretical side,   Lun et al.\ \cite{LSJC84} have obtained the rheological properties of a dense gas for small inelasticity, while  Jenkins and Richman \cite{JR88} have used a maximum-entropy approximation to solve the Enskog equation. This method  has been subsequently extended to highly inelastic spheres \cite{CR98}. Sela et al.\ \cite{SGN96} have solved the  Boltzmann equation to third order in the shear rate, finding normal stress differences. Some progress has been made by  using model kinetic equations for dilute granular gases  \cite{brey97shear}, as well as for dense granular gases  \cite{MGSB99}. Exact solutions derived in both cases compare quite well with Monte Carlo simulations, even for strong  dissipation. Similar studies for multi-component systems are much scarcer, although some recent work has been  carried  out \cite{AL03,CH02,MG02,MG03}. }   

 Normally, transport properties are defined for general states with the temperature, shear rate, and restitution coefficient as independent variables. For example, the shear viscosity as given by the Chapman--Enskog solution to the Boltzmann equation \cite{BDKyS98} has the form for shear flow $\eta =\eta \left(T, a,\alpha \right) $, and the Newtonian shear viscosity is obtained from it at zero shear rate, $\eta _{0}\left(T, \alpha\right) =\eta \left(T, a=0,\alpha \right)$.  However, in the steady state of USF the condition for balancing collisional cooling and viscous heating implies that the steady state temperature is a function of the shear rate $a$ and the coefficient of restitution $\alpha $ measuring the degree of inelasticity of the particles, $T_{s}=T_{s}(a,\alpha )$. In  this case $\eta =\eta \left(T_{s}(a,\alpha ), a,\alpha\right) =\eta _{s}(a,\alpha )=\eta _{s}^{\prime }(T_{s},\alpha )$. {This complication raises two interesting questions: }  
\begin{enumerate} 
\item 
{For sufficiently small }$a${, is it possible to describe USF using Newtonian hydrodynamics?}  
\item 
{If }$\eta _{s}(a,\alpha )$ { is measured in USF, can the Newtonian viscosity }$\eta _{0}\left( T,\alpha \right) ${ be deduced from it? } 
\end{enumerate}   
{ The objective of this paper is to answer these two questions in the general context of hydrodynamics for granular fluids. It is shown that the answer is negative in both cases, implying that USF is inherently non-Newtonian and that the full nonlinear dependence of }$\eta \left( T,a,\alpha \right) ${ on the shear rate is required, even for small }$a${, at any }$\alpha \neq 1${. This result is important because it identifies a fundamental and unbridgeable gap between Navier--Stokes hydrodynamics and the hydrodynamics for steady shear flow. There is no possibility of using USF to study the Newtonian viscosity (e.g., by molecular dynamics simulation); there is no possibility of using the Navier--Stokes equations to describe USF or states near it (e.g., the stability of USF). This is not an entirely negative result, however, because it shows that USF is a rich testing ground for the study of rheology. Any measurement will necessarily be illustrating some rheological effect.} 
   
{These results may seem counter-intuitive at first as with many conceptual differences between granular and real fluids. However, the steady state considered here for the granular fluid does not exist for a real fluid so intuition is not reliable in this case. The analysis presented here, and illustrated in detail for a dilute gas, is important because similar failures of the Navier--Stokes level hydrodynamics must be expected for other steady states specific to granular fluids. }This peculiar limitation of Newtonian hydrodynamics may apply  generically when the steady state exists only because of the internal cooling mechanism, rather than as a consequence of controllable external fields or boundary forces. The steady state implies the equivalence of spatial heat conduction, viscous heating, and collisional cooling, and a consequent relationship between the heat flux, momentum flux, and cooling rate (see below). For a given cooling rate (i.e., given restitution coefficient) the gradients of the hydrodynamic fields can no longer be controlled independently to assure the validity of the Newtonian limit.      
This phenomenon is elaborated further in the next section for USF based on the macroscopic balance equations for mass, energy, and momentum, and idealized boundary conditions for a general granular fluid. {This analysis assumes only the existence of a hydrodynamic description, but is not limited to any small gradient, small inelasticity, or other approximation.} {A more detailed illustration from the Boltzmann kinetic theory  is given in Section \ref{sec3}}. {An important point of this kinetic theory analysis is the determination of the rheological equation of state }$ \eta \left( T,a,\alpha \right) ${ for} \emph{both} steady and { unsteady} states of USF. In the {unsteady} state the temperature increases (decreases) depending on whether the viscous heating of the initial state is larger (smaller) than the collisional cooling. The three parameters $ a,\alpha ,T$ can be controlled independently, and there is a small shear rate domain for which the {unsteady} fluid is Newtonian. In the steady state, {however,} these parameters are constrained, $T=T_{s}(a,\alpha )$, such that $\eta \left( T_{s}(a,\alpha ),a,\alpha \right) $ is never in its Newtonian limit for any $\alpha <1$. These results {and their importance for the study of transport in granular fluids} are summarized with comments in the last section.    \section{Hydrodynamics for USF\label{sec2}}    An idealized granular fluid consists of smooth hard spheres ($d=3$) or disks ($d=2$) of diameter $\sigma $ and mass $m$. The collisions between particles are characterized through a constant coefficient of normal restitution $ \alpha $ with values $0<\alpha \leq 1$. The elastic limit corresponds to $ \alpha =1$. The exact macroscopic balance equations for mass, energy, and momentum are  \begin{equation} D_{t}n+n\nabla \cdot \mathbf{u}=0,  \label{2.1} \end{equation} \begin{equation} \left( D_{t}+\zeta \right) T+\frac{2}{dn}\left( P_{ij}\nabla _{j}u_{i}+\nabla \cdot \mathbf{q}\right) =0,  \label{2.2} \end{equation} \begin{equation} D_{t}u_{i}+(mn)^{-1}\nabla _{j}P_{ij}=0,  \label{2.3} \end{equation} where $n$ is the density, $T$ is the granular temperature, $\mathbf{u}$ is the flow velocity,  $D_{t}=\partial _{t}+\mathbf{u}\cdot \nabla $ is the material derivative, { $\zeta$ is the cooling rate,   $\mathsf{P}$  is the pressure tensor, and $\mathbf{q}$  is the heat flux. On physical grounds, $\zeta$ has essentially the form  $\zeta\approx \nu(1-\alpha^2)$, where $\nu\propto \sqrt{T}$ is a mean collision frequency and $1-\alpha^2$ is the  fraction of energy lost in each inelastic collision. In order to get a closed set of hydrodynamic equations from  (\ref{2.1})--(\ref{2.3}),   $\zeta$, $\mathsf{P}$, and $\mathbf{q}$  must be specified as functionals of the fields $n$, $T$, and $\mathbf{u}$.}  However, some interesting results can be obtained for the purposes here even at this exact level. {We emphasize that the analysis of this section and its conclusion do not depend on any specific form for the rheological equation of state, only its scaling with respect to the available hydrodynamic fields (dimensional analysis).}    

An idealized macroscopic state of USF is characterized by the forms \cite{GS03} \begin{equation} n(\mathbf{r},t)=n,\quad T(\mathbf{r},t)=T(t),\quad  u_{x}=ay,\quad  u_{y}=u_{z}=0.  \label{2.4} \end{equation} The presumed geometry is that described above. More precisely, this linear velocity profile assumes no boundary layer near the walls and is possible for special periodic boundary conditions in the local Lagrangian frame  \cite{Lees}. Since the density is a constant, it plays no significant role in the following and the dependence of properties on it will be suppressed. From symmetry, the cooling rate, heat flux, and  pressure tensor must have the forms in the hydrodynamic state  \begin{equation} \zeta =\zeta \left( T(t),a,\alpha \right) ,\quad \mathbf{q=0}, \quad \nabla _{j}P_{ij}=0  .\label{2.5} \end{equation}    The balance equations (\ref{2.1}) and (\ref{2.3}) are satisfied exactly with these choices, while Eq.\ (\ref{2.2}) becomes  \begin{equation} \partial _{t}T(t)=-\frac{2}{dn}aP_{xy}\left( T(t),a,\alpha \right) -\zeta \left( T(t),a,\alpha \right) T(t).  \label{2.8} \end{equation} The functional forms for the scalars $\zeta \left( T,a,\alpha \right) $ and $ P_{xy}\left( T,a,\alpha \right) $ must be determined from a more microscopic basis. An example from kinetic theory is provided in the next section. Then, Eq.\ (\ref{2.8})  provides a closed hydrodynamic equation to determine the temperature $T(t)$ for given initial condition $T(0)$ and specification of the constants $a$ and $\alpha $. The first term on the right side is positive and represents viscous heating. To make this more explicit it is usual to introduce the nonlinear shear viscosity $\eta \left( T,a,\alpha \right) $ by  \begin{equation} P_{xy}\left( T,a,\alpha \right) \equiv -\eta \left( T,a,\alpha \right) a. \label{2.9} \end{equation} A Newtonian fluid is that for which the viscosity and cooling rate becomes independent of the shear rate,  \begin{eqnarray} &&\eta \left( T,a,\alpha \right) \rightarrow \eta \left( T,a=0,\alpha \right) \equiv \eta _{0}\left( T,\alpha \right) ,\nn && \zeta \left( T,a,\alpha \right) \rightarrow \zeta \left( T,a=0,\alpha \right) \equiv \zeta _{0}\left( T,\alpha \right) . \label{2.10} \end{eqnarray} The corresponding hydrodynamic equation for a Newtonian fluid becomes  \beqa \partial _{t}T(t)&=&-\zeta _{0}\left( T(t),\alpha \right) T(t)\nn &&+a^{2}\left[ \frac{2}{dn}\eta _{0}\left( T(t),\alpha \right) -\zeta _{2}\left( T(t),\alpha \right) T(t)\right] .  \label{2.11} \eeqa Here, for consistency, the cooling rate also has been expanded to second order in the shear rate,
\beq 
\zeta \left( T(t),a,\alpha \right) \rightarrow \zeta _{0}\left( T(t),\alpha \right) +a^{2}\zeta _{2}\left( T(t),\alpha \right) . 
\label{2.12} 
\eeq 
The Newtonian fluid equation (\ref{2.11}) appears quite useful as a means to measure its viscosity. First, the cooling rate $\zeta _{0}\left( T(t),\alpha \right) $ is determined by measuring the time dependence of $T(t)$ at different values of $\alpha $ at zero shear rate. Next, the same measurement is performed as a function of the shear rate to determine the combination $\left( 2/dn\right) \eta _{0}\left( T(t),\alpha \right) -\zeta _{2}\left( T(t),\alpha \right) T(t)$. In the case $\zeta _{2}=0$ (see the next section for an example) this determines the Newtonian viscosity.    Two questions arise at this point. First, can the measurement of the Newtonian viscosity be simplified by considering only the steady state? Second, under what conditions does the Newtonian limit for the steady state apply?  To answer these questions we have to take into account that the domain of validity of the Newtonian description is  restricted to small shear rates. To be more precise, a dimensionless shear rate must be introduced. The only other relevant frequency for the fluid is its mean collision frequency $\nu (T(t))\propto\sqrt{T(t)}$. Consequently, the relevant dimensionless shear rate, cooling rate, and shear viscosity are defined by  \beqa &&a^{\ast }(T(t))=\frac{a}{\nu (T(t))}, \quad  \zeta ^{\ast }\left(a^{\ast },\alpha \right) =\frac{\zeta \left( T(t),a,\alpha \right) }{\nu (T(t))},\nn && \eta ^{\ast }\left( a^{\ast },\alpha \right) =\frac{\nu (T(t))  }{nT(t)}  \eta \left( T(t),a,\alpha \right),\label{2.14}   
\eeqa   
where $\zeta ^{\ast }\left( a^{\ast },\alpha \right) $ and $\eta ^{\ast   
}\left( a^{\ast },\alpha \right) $  can depend on time and temperature only   
through $a^{\ast }(T(t))$, in order to be dimensionless. { Note that  the reduced shear rate $a^*$ is actually a    
measure of the granular temperature since $T\propto {a^*}^{-2}$. In the steady   
state, Eq.\  (\ref{2.8}) becomes    
\begin{equation}   
a^{\ast 2}=\frac{d\zeta ^{\ast }\left( a^{\ast },\alpha \right) }{2\eta   
^{\ast }\left( a^{\ast },\alpha \right) }.  \label{2.15}   
\end{equation}   
The solution to this equation gives $a_s^*(\alpha)$, that   
depends only on $\alpha $ and   
thus cannot be made small by controlling the shear rate $a$.    
 In general, $a^{\ast }_s(\alpha)\ll 1$ only in the quasielastic  limit $1-\alpha \ll 1$,   
\beq   
{a_s^*}^2(\alpha)\to \frac{d\zeta_0^*(\alpha)}{2\eta_0^*(\alpha=1)}\equiv {a_0^*}^2(\alpha)\sim 1-\alpha\ll 1,   
\label{2.15bis}   
\eeq   
{ where we have taken into account that $\zeta_0^*(\alpha)\propto 1-\alpha^2\sim 2(1-\alpha)$ in the quasielastic    
limit.}   
 In this case   
the dimensionless steady state shear viscosity becomes    
$\eta_s^*(\alpha)=\eta^*(a^*_s(\alpha),\alpha)\to\eta^*(a_0^*(\alpha),\alpha)$. Since, according to Eq.\    
(\ref{2.15bis}), $a_0^*$ is small, the steady state shear viscosity is expected to differ from the dimensionless    
Newtonian viscosity $\eta_0^*(\alpha)=\eta^*(a^*=0,\alpha)$ by a (super-Burnett) term of order ${a_0^*}^2$:}   
\beq   
\eta^*_s(\alpha)-\eta_0^*(\alpha)\sim {a_0^*}^2(\alpha)\sim 1-\alpha.   
\label{2.16a}   
\eeq   
On the other hand, { by expanding the Newtonian shear viscosity $\eta_0^*(\alpha)$ around its value in the elastic    
case, we have   
\beq   
\eta_0^*(\alpha)-\eta_0^*(\alpha=1)\sim  1-\alpha,   
\label{2.16b}   
\eeq   
for small inelasticities.}   
Therefore, {Eqs.\ (\ref{2.16a}) and (\ref{2.16b}) show that} in the quasielastic limit the steady state shear    
viscosity $\eta^*_s(\alpha)$ differs from the Newtonian viscosity $\eta_0^*(\alpha)$ as much as the latter differs    
from that of the elastic gas.   
This shows that a measurement of the steady state temperature to determine $a_s^{\ast }(\alpha)$ does not allow    
determination of the Newtonian viscosity, even for    
$1-\alpha \ll 1$. The $\alpha $ dependence of $\eta_0^*(\alpha)=\eta ^{\ast }\left( a^{\ast   
}=0,\alpha \right) $ cannot be isolated from the $\alpha $ dependence of $\eta ^{\ast }\left( a_{s}^{\ast }(\alpha    
),\alpha \right) $. Both viscosities coincide only in the trivial case $\alpha=1$.

In summary, for any chosen $a,\alpha ,T(0)$ the steady state viscosity is $\eta ^{\ast }\left( a_s^{\ast }(\alpha    
),\alpha \right) $ and the system   
samples a non-Newtonian value from the general rheological equation of state    
$\eta ^{\ast }\left( a^{\ast },\alpha \right) $ that is always different   
from the Newtonian value.

\section{{ Description from the Boltzmann kinetic theory\label{sec3}}}   
To explore this phenomenon in more detail it is necessary to calculate the   
rheological equation of state $\eta ^{\ast }\left( a^{\ast },\alpha \right)$.    
This requires a more microscopic analysis such as kinetic theory. At low   
density the granular Boltzmann equation provides the appropriate starting   
point.    
   
\subsection{Boltzmann equation and Newtonian viscosity}   
The Boltzmann kinetic equation determines the probability density $f(\mathbf{   
r},\mathbf{v},t)$ for a particle to have position $\mathbf{r}$ and velocity $   
\mathbf{v}$ at time $t$ in a low density gas. It has the form    
\begin{equation}   
\left( \frac{\partial }{\partial t}+\mathbf{v}\cdot \nabla \right) f(\mathbf{   
r},\mathbf{v},t)=J[\mathbf{r},\mathbf{v}|f(t)].  \label{a.1}   
\end{equation}   
The right side describes the effects of inelastic pair collisions. The   
detailed form of the Boltzmann collision operator, $J$, is not required here   
beyond noting the properties necessary for the macroscopic balance equations    
\begin{equation}   
\int d\mathbf{v}\left(    
\begin{array}{c}   
1 \\    
\mathbf{v} \\    
\frac{1}{2}m\left( \mathbf{v}-\mathbf{u}\right) ^{2}   
\end{array}   
\right) J[\mathbf{r},\mathbf{v}|f(t)]=\left(    
\begin{array}{c}   
0 \\    
\mathbf{0} \\    
-\frac{d}{2}nT\zeta   
\end{array}   
\right) .  \label{a.2}   
\end{equation}   
Here  the density $n$,  the temperature $T$, and  the   
macroscopic flow velocity $\mathbf{u}$ are defined in terms of $f(\mathbf{r},\mathbf{v},t)$   
by    
\begin{equation}   
\left(    
\begin{array}{c}   
n(\mathbf{r},t) \\    
n(\mathbf{r},t)\mathbf{u}(\mathbf{r},t) \\    
\frac{d}{2}n(\mathbf{r},t)T(\mathbf{r},t)   
\end{array}   
\right) =\int d\mathbf{v}\left(    
\begin{array}{c}   
1 \\    
\mathbf{v} \\    
\frac{1}{2}m\left( \mathbf{v}-\mathbf{u}\right) ^{2}   
\end{array}   
\right) f(\mathbf{r},\mathbf{v,}t).  \label{a.3}   
\end{equation}   
The two zeros on the right side of Eq.\ (\ref{a.2}) correspond to conservation of   
mass and momentum, while the last term results from non-conservation of energy.   
The properties (\ref{a.2}) lead to the macroscopic balance equations (\ref{2.1})--(\ref{2.3}), with    
the following microscopic expressions for the pressure tensor, the heat flux, and the cooling rate:   
\beq   
P_{ij}(\mathbf{r},t)=m\int d\mathbf{v} V_iV_j f(\mathbf{r},\mathbf{v},t),   
\label{n1}   
\eeq   
\beq   
\mathbf{q}(\mathbf{r},t)=\frac{m}{2}\int d\mathbf{v} V^2\mathbf{V} f(\mathbf{r},\mathbf{v},t),   
\label{n2}   
\eeq   
\beqa   
&&\zeta(\mathbf{r},t) =(1-\alpha ^{2})\frac{m\pi ^{\frac{d-1}{2}}\sigma ^{d-1}}{4d\Gamma   
\left( \frac{d+3}{2}\right) n(\mathbf{r},t)T(\mathbf{r},t)}\nonumber\\   
&&\times\int d{\bf v}\int d{\bf v}_{1}\,|{\bf v   
}-{\bf v}_{1}|^{3}f({\bf r},{\bf v},t)f({\bf r},{\bf v}_{1},t).  \label{3.5.2}   
\eeqa   
In Eqs.\ (\ref{n1}) and (\ref{n2}), $\mathbf{V}=\mathbf{v}-\mathbf{u}(\mathbf{r},t)$ is the peculiar velocity.    
   
It is straightforward to determine the Navier--Stokes shear viscosity coefficient $\eta_0(T,\alpha)$ by using the    
Chapman--Enskog method \cite{CC70}. The result is \cite{BDKyS98,GD99,BC01}   
\beq   
\eta_0(T,\alpha)=\frac{nT}{\nu(T)}\eta_0^*(\alpha),   
\label{n3}   
\eeq   
where    
\begin{equation}   
\nu(T) =\frac{8\pi ^{(d-1)/2}\sigma ^{d-1}}{(d+2)\Gamma \left( d/2\right) }   
n\left( \frac{T}{m}\right) ^{1/2}.  \label{a.7}   
\end{equation}   
is an effective collision frequency, and    
\begin{equation}   
\eta _{0}^{\ast }\left( \alpha \right) =\left[ \beta \left( \alpha \right) +   
\frac{1}{2}\zeta _{0}^{\ast }\left( \alpha \right) \right] ^{-1}.   
\label{3.1}   
\end{equation}   
Here $\zeta_0^*(\alpha)=\zeta_0/\nu$ is the  dimensionless cooling rate    
in the homogeneous cooling state, while $\beta \left( \alpha \right) $ is a dimensionless function of the   
restitution coefficient given in terms of the solution to the linearized Boltzmann equation. Explicit results for    
$\zeta_0^*$ and $\beta$  can be obtained by considering the leading terms in a Sonine polynomial expansion. In that    
approximation,   
\beq   
 \zeta_0^*(\alpha)=\frac{d+2}{4d}(1-\alpha^2) ,  \label{3.2b}   
\eeq   
\beq   
\beta \left( \alpha \right) =\frac{1+\alpha }{2}\left[ 1-\frac{d-1}{2d}   
(1-\alpha )\right]. \label{3.2a}   
\eeq   
   
\subsection{Uniform shear flow}   
It becomes prohibitively difficult to go beyond Navier--Stokes order to get   
the {exact} full dependence of $\eta ^{\ast }\left( a^{\ast },\alpha \right) $ on $a^{\ast }$ from the Boltzmann    
equation.    
{On the other hand, a good estimate is provided by considering the leading Sonine approximation to the distribution    
function (Grad's method) \cite{MG02,G02}. 
   
In the special case of USF the solution to the Boltzmann kinetic equation is spatially   
uniform when expressed in terms of the velocity relative to the local flow, $V_{x}=v_{x}-ay,V_{y}=v_{y},V_{z}=v_{z}$.    
Consequently, the Boltzmann  equation (\ref{a.1}) becomes    
\cite{GS03}    
\begin{equation}   
\left( \partial _{t}-aV_{y}\frac{\partial }{\partial V_{x}}\right) f(\mathbf{V},t)=J[\mathbf{V}|f(t)].  \label{a.11}   
\end{equation}   
Multiplying both sides of Eq.\ (\ref{a.11}) by $mV_{i}V_{j}$ and integrating   
over velocity, we get    
\begin{eqnarray}   
\partial _{t}P_{ij}+a\left( \delta _{ix}P_{yj}+\delta _{jx}P_{iy}\right)   
&=&m\int d\mathbf{v} V_i V_j J[\mathbf{V}|f]\nonumber\\   
&\equiv&-\Lambda_{ij}.   
\label{a.12}   
\end{eqnarray}   
The exact expression of the collision integral $\Lambda_{ij}$ is not known, even in the elastic case. However, a good    
estimate can be expected by using  Grad's approximation   
\beq   
f(\mathbf{V})\to f_0(\mathbf{V})\left[1+\frac{m}{2T}\left(\frac{P_{ij}}{nT}-\delta_{ij}\right)V_iV_j\right],   
\label{n4}   
\eeq   
where    
\begin{equation}   
f_{0}(\mathbf{V})=n(m/2\pi T)^{d/2}\exp (-mV^{2}/2T)  \label{a.6}   
\end{equation}   
is the local equilibrium distribution function.   
When Eq.\ (\ref{n4}) is inserted into the definition of $\Lambda_{ij}$ and terms nonlinear in $P_{ij}/nT-\delta_{ij}$    
are neglected, one gets \cite{G02}   
\beq   
\Lambda_{ij}=\nu\left[\beta \left( P_{ij}-\delta _{ij}\right) +\zeta _{0}^*P_{ij}\right],   
\label{n5}   
\eeq   
where $\zeta_0^*$ and $\beta$ are given by Eqs.\ (\ref{3.2b}) and (\ref{3.2a}), respectively.   
{It is worth noting that Eq.\ (\ref{n5}) coincides with the one obtained from a simple kinetic model    
\cite{brey97shear,brey99model3}.}       
   
The three relevant independent equations from (\ref{a.12}) and (\ref{n5}) are    
\begin{equation}   
\partial _{t}p+\zeta _{0}p+\frac{2a}{d}P_{xy}=0,  \label{a.14}   
\end{equation}   
\begin{equation}   
\partial _{t}P_{xy}+\left( \beta \nu+\zeta _{0}\right) P_{xy}+aP_{yy}=0,   
\label{a.15}   
\end{equation}   
\begin{equation}   
\partial _{t}P_{yy}+\left( \beta \nu+\zeta _{0}\right) P_{yy}-\beta \nu   
p=0.  \label{a.16}   
\end{equation}   
where $p=nT=P_{ii}/d$ is the low density pressure. If we define the   
dimensionless quantities    
\beqa   
P_{ij}^{\ast }(t)&=&\frac{P_{ij}(t)}{nT(t)},\quad  a^{\ast }(t)=\frac{a}{\nu (T(t))},\nn   
  \tau(t)&=&\int_{0}^{t}dt^\prime \nu (T(t^{\prime })),  \label{a.17}   
\eeqa   
then Eqs.\ (\ref{a.14})--(\ref{a.16}) become    
\begin{equation}   
2\partial _{\tau}\ln a^*=\zeta _{0}^{\ast }+\frac{2}{d}P_{xy}^{\ast }a{^{\ast }},   
\label{3.3}   
\end{equation}   
\begin{equation}   
\partial _{\tau}P_{xy}^{\ast }=-a^{\ast }P_{yy}^{\ast }-P_{xy}^{\ast }\left(   
\beta -\frac{2}{d}P_{xy}^{\ast }a{^{\ast }}\right) ,  \label{3.4}   
\end{equation}   
\begin{equation}   
\partial _{\tau}P_{yy}^{\ast }=-P_{yy}^{\ast }\left( \beta -\frac{2}{d}   
P_{xy}^{\ast }a{^{\ast }}\right) +\beta .  \label{3.5}   
\end{equation}   
The variable $\tau$ is the dimensionless time  measured as the average collision number.   
The solution to these nonlinear equations gives    
{ $a^*$, $P_{xy}^{\ast }$, and $P_{yy}^{\ast }$ as functions of $\tau$ for a given value of $\alpha$. The   
rheological equation of state is then obtained from    
\begin{equation}   
\eta ^{\ast }(a^*(\tau))=-\frac{P_{xy}^{\ast }(\tau)}{a^{\ast   
}(\tau)}.  \label{3.6}   
\end{equation}

\subsubsection{Steady state solution}   
Consider first the steady state solution. Equation (\ref{3.3}) gives    
\begin{equation}   
P_{xy,s}^{\ast }=-\frac{d\zeta _{0}^{\ast }}{2a_s^{\ast }},  \label{3.7}   
\end{equation}   
while Eqs.\  (\ref{3.4}) and (\ref{3.5}) give    
\begin{equation}   
P_{yy,s}^*={\beta}\left({\beta+\zeta_0^*}\right)^{-1},\quad P{_{xy,s}^{\ast }=}-a_s^{\ast }\beta \left( \beta +\zeta    
_{0}^{\ast }\right)   
^{-2}.  \label{3.7a}   
\end{equation}   
The value of    
$a^{\ast }$ in the steady state is obtained from Eqs.\ (\ref{3.7}) and (\ref{3.7a}),    
\begin{equation}   
a_{s}^{\ast }(\alpha)=\sqrt{\frac{d\zeta _{0}^{\ast }(\alpha)}{2\beta \left( \alpha \right) }   
}\left[ \beta \left( \alpha \right) +\zeta _{0}^{\ast }\left( \alpha \right)   
\right] .  \label{3.8a}   
\end{equation}   
As anticipated above, it is independent of the initial temperature and shear   
rate $a$.    
Appendix \ref{appA} shows that the steady state solution (\ref{3.7a})--(\ref{3.8a}) is indeed a (linearly) stable    
solution.

The second equality in (\ref{3.7a}) allows one to identify the steady state shear viscosity as    
\begin{equation}   
\eta _{s}^{\ast }\left( \alpha \right) =-\frac{P_{xy}^{\ast }}{a^{\ast }}   
=\beta \left( \alpha \right) \left[ \beta \left( \alpha \right) +\zeta   
_{0}^{\ast }\left( \alpha \right) \right] ^{-2}.  \label{3.8}   
\end{equation}   
This gives the explicit form for the { steady state} shear viscosity, showing it is also   
independent of  {the} shear rate $a$ and the initial temperature.   
Furthermore, its dependence on $\alpha $ is qualitatively different from   
that of Eq.\ (\ref{3.1}) for a Newtonian fluid. This is illustrated in Fig.\ \ref{fig1}, which also shows that the analytical results    
compare favorably well with simulation data obtained from the DSMC method \cite{MSG04,AS04}. Clearly, there is  no    
relationship of the steady state shear   
viscosity to the Newtonian viscosity at any value of $\alpha $.    
{While the Newtonian viscosity is $\eta_0^*(\alpha)>1$, the steady state viscosity is $\eta_s^*(\alpha)<1$. In the    
quasielastic limit, Eqs.\ (\ref{3.1}), (\ref{3.8a}), and (\ref{3.8}) yield   
\beq   
\eta_0^*(\alpha)\to 1+\frac{3d-4}{4d}(1-\alpha),   
\label{3new1a}\eeq   
\beq   
{a_s^*}^2(\alpha)\to {a_0^*}^2(\alpha)=\frac{{d+2}}{4}\left(1-\alpha\right),   
\label{3new2}   
\eeq   
\beq   
 \eta_s^*(\alpha)\to 1-\frac{5}{2d}(1-\alpha),   
\label{3new1b}   
\eeq   
}{ Equations (\ref{3new1a})--(\ref{3new1b}) confirm the qualitative arguments behind Eqs.\    
(\ref{2.15bis})--(\ref{2.16b}).}   
Eliminating $\alpha$ between Eqs.\ (\ref{3new2}) and (\ref{3new1b}), one has    
\beq   
\eta_s^*(\alpha)\to 1-\frac{10}{d(d+2)}{a_s^*}^2(\alpha)   
\label{n10}   
\eeq   
in the quasielastic limit.   
\begin{figure}[tbp]   
\includegraphics[width=.90 \columnwidth]{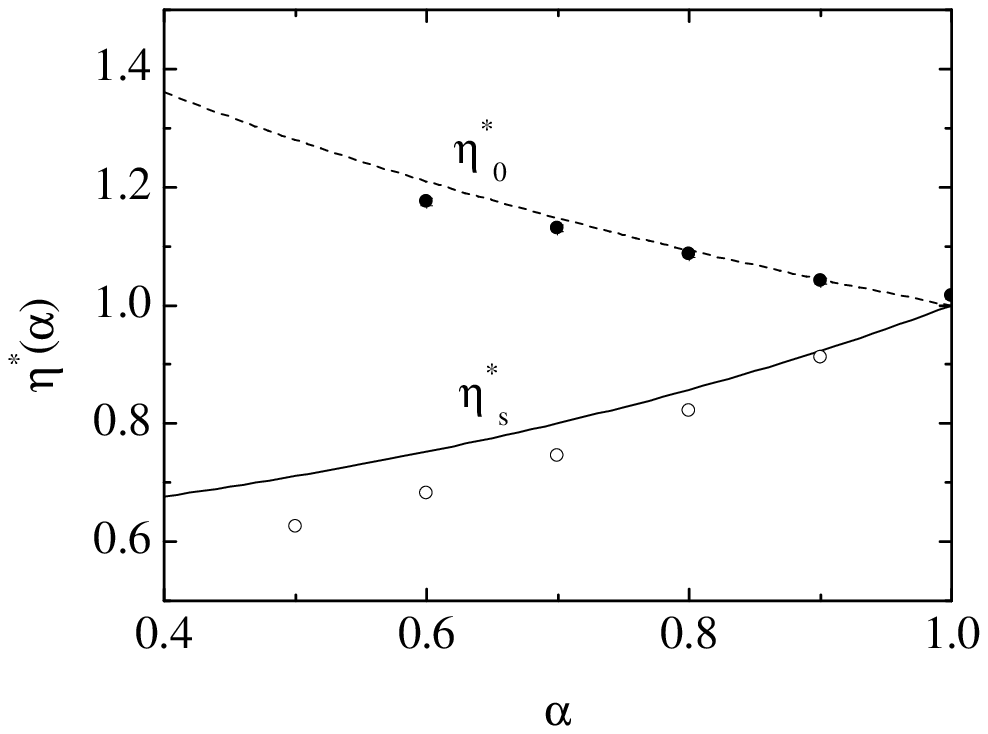}   
\caption{{ Plot of the dimensionless Newtonian shear viscosity $\eta _{0}^{*}(\alpha)$ (dashed line) and the    
dimensionless steady state shear viscosity $\eta_{s}^{*}(\alpha)$ (solid line) for $d=3$}, { as given by Eqs.\    
(\protect\ref{3.1}) and (\protect\ref{3.8}), respectively. Filled circles \protect\cite{MSG04} and open circles    
\protect\cite{AS04}  represent simulation data obtained from the numerical solution of the Boltzmann equation by the    
DSMC method.}    
\label{fig1}}   
\end{figure}   
   
\subsubsection{Unsteady hydrodynamic solution}   
   
The more general solution to Eqs.\ (\ref{3.3})--(\ref{3.5}) corresponding to   
hydrodynamics is that for which all time dependence occurs through the   
hydrodynamic fields (the Chapman--Enskog ``normal'' solution). The time   
dependence is due only to the temperature for USF which occurs in the   
dimensionless forms above through their dependence { on $a^{\ast }(t)=a/\nu   
(T(t))$},   
\begin{equation}   
\partial _{\tau}P_{ij}^{\ast }=\frac{\partial P_{ij}^{\ast }}{\partial a^{\ast   
}}\partial _{\tau} a^*.  \label{3.9}   
\end{equation}   
Then, using Eq.\ (\ref{3.3}), Eqs.\ (\ref{3.4})   
and (\ref{3.5}) become    
\begin{equation}   
\frac{\partial P_{xy}^{\ast }}{\partial a^{\ast }}=\frac{-2P_{yy}^{\ast }-   
\frac{2}{a^{\ast }}P_{xy}^{\ast }\left( \beta -\frac{2}{d}P_{xy}^{\ast }a{   
^{\ast }}\right) }{ \zeta _{0}^{\ast }+\frac{2}{d}P_{xy}^{\ast }a{   
^{\ast }} },  \label{3.10}   
\end{equation}   
\begin{equation}   
\frac{\partial P_{yy}^{\ast }}{\partial a^{\ast }}=\frac{2\beta   
-2P_{yy}^{\ast }\left( \beta -\frac{2}{d}P_{xy}^{\ast }a{^{\ast }}\right) }{   
a^{\ast }\left( \zeta _{0}^{\ast }+\frac{2}{d}P_{xy}^{\ast }a{^{\ast }}   
\right) }.  \label{3.11}   
\end{equation}   
This is a set of two coupled nonlinear differential equations that must be 
solved with the appropriate boundary conditions to get the \textit{ 
hydrodynamic} solution. There is a singular point corresponding to the 
steady state solution (\ref{3.7a}) and (\ref{3.8a}), in which case the 
numerators and denominators of Eqs.\ (\ref{3.10}) and (\ref{3.11}) vanish.

The { unsteady} hydrodynamic solutions $\eta ^{\ast }\left( a^{\ast },\alpha \right) 
=-P_{xy}^{\ast }/a^{\ast }$ are illustrated in Fig.\ \ref{fig2} for three 
different values of $\alpha $. Also shown are the special values of the   
steady state shear rate $a_s^*(\alpha)$ and shear viscosity $\eta_s^*(\alpha)$ for each curve.    
For each value of $\alpha$ the point $(a_s^*,\eta_s^*)$ splits the curve $\eta^*(a^*)$ into two physically different    
branches, one for $a^*<a_s^*$ and another one for $a^*>a_s^*$.   
Suppose that for a given shear   
rate $a$ and initial temperature $T(0)$  the value of $a^{\ast }$ is less than that for the steady state,    
$a^*(0)=a/\nu(T(0))<a^*_s$.    
The cooling dominates viscous heating in this case and the temperature   
decreases, leading to a larger value of $a^{\ast }(\tau)$. In this way, the system   
evolves according to the hydrodynamic equations along the curve until the   
steady state value  $a_s^{\ast }$ is attained, i.e., $a^*(\tau)\to a_s^*$ as $\tau\to \infty$.    
A parametric plot of $\eta^*(\tau)$ versus $a^*(\tau)$ gives the branch of the curve $\eta^*(a^*)$ corresponding to    
$a^*< a_s^*$.   
Analogously,    
a heating process   
occurs in the opposite case of an initial value for $a^{\ast }$ greater than   
that for the steady state until  $\lim_{\tau\to\infty}a^*(\tau)=a_s^*$ again. This provides the branch corresponding    
to $a^*> a_s^*$.   
As an example, Fig.\ \ref{fig4} shows the time evolution of $a^*(\tau)\propto 1/\sqrt{T(\tau)}$, $-P_{xy}^*(\tau)$,    
and $P_{yy}^*(\tau)$ for $\alpha=0.5$ and two different initial conditions: (i) $a^*(0)=0.1$, $P_{xy}^*(0)=-0.1$,     
$P_{yy}^*(0)=1$, and (ii) $a^*(0)=2$, $P_{xy}^*(0)=-0.5$,  $P_{yy}^*(0)=0.4$. In both cases, after about 20--30    
collisions per particle, the system reaches a common steady state with $a^*_s=0.812$, $-P_{xy,s}^*=0.577$, and    
$P_{yy,s}^*=0.667$. We have checked that the same asymptotic state is achieved when starting from different initial    
conditions.   
\begin{figure}[tbp]   
\includegraphics[width=.90 \columnwidth]{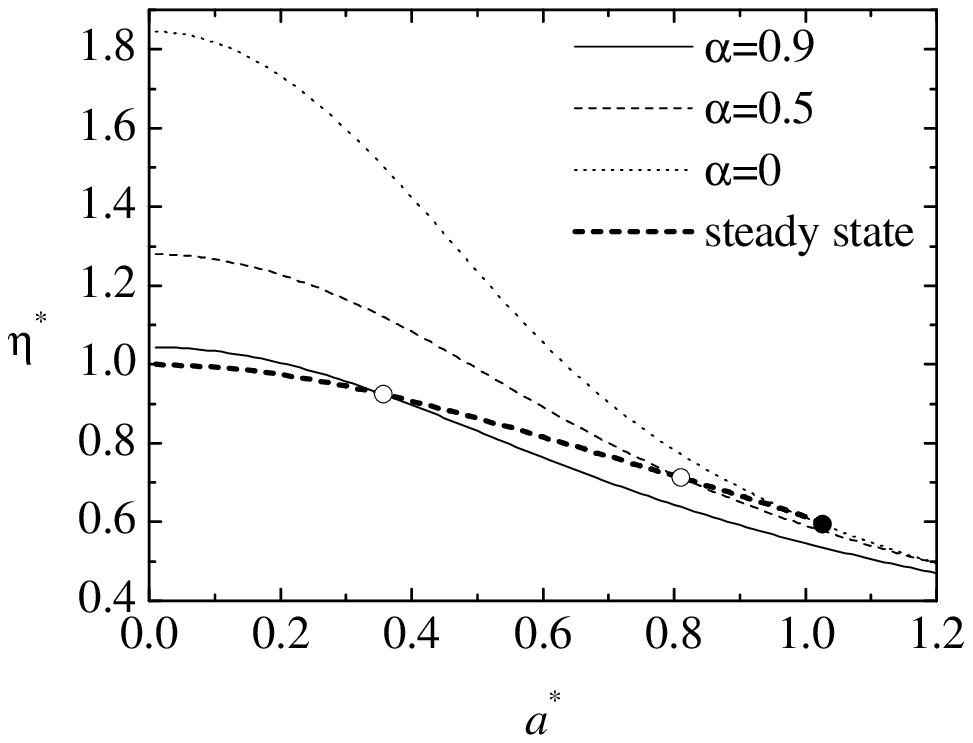}   
\caption{Plot of $\eta ^{\ast }(a^{\ast },\alpha )$ as a   
function of $a^{\ast }$ for $d=3$ and $\alpha =0.9$ (solid line), $\alpha =0.5$ (dashed line), and $\alpha =0$ (dotted    
line). The thick   
dashed line is the locus of points $(a_{s}^{\ast },\eta _{s}^{\ast })$, { which are parametrically found from Eqs.\    
(\protect\ref{3.8a}) and (\protect\ref{3.8})}. It intercepts the curves representing $\eta ^{\ast}(a^{\ast },\alpha )$    
at the steady state values (indicated by   
circles). { Note that the curve $(a_{s}^{\ast },\eta _{s}^{\ast })$ ends at the point corresponding to $\alpha=0$    
(represented by a filled circle)}.   
\label{fig2}}   
\end{figure}   
\begin{figure}[tbp]   
\includegraphics[width=.90 \columnwidth]{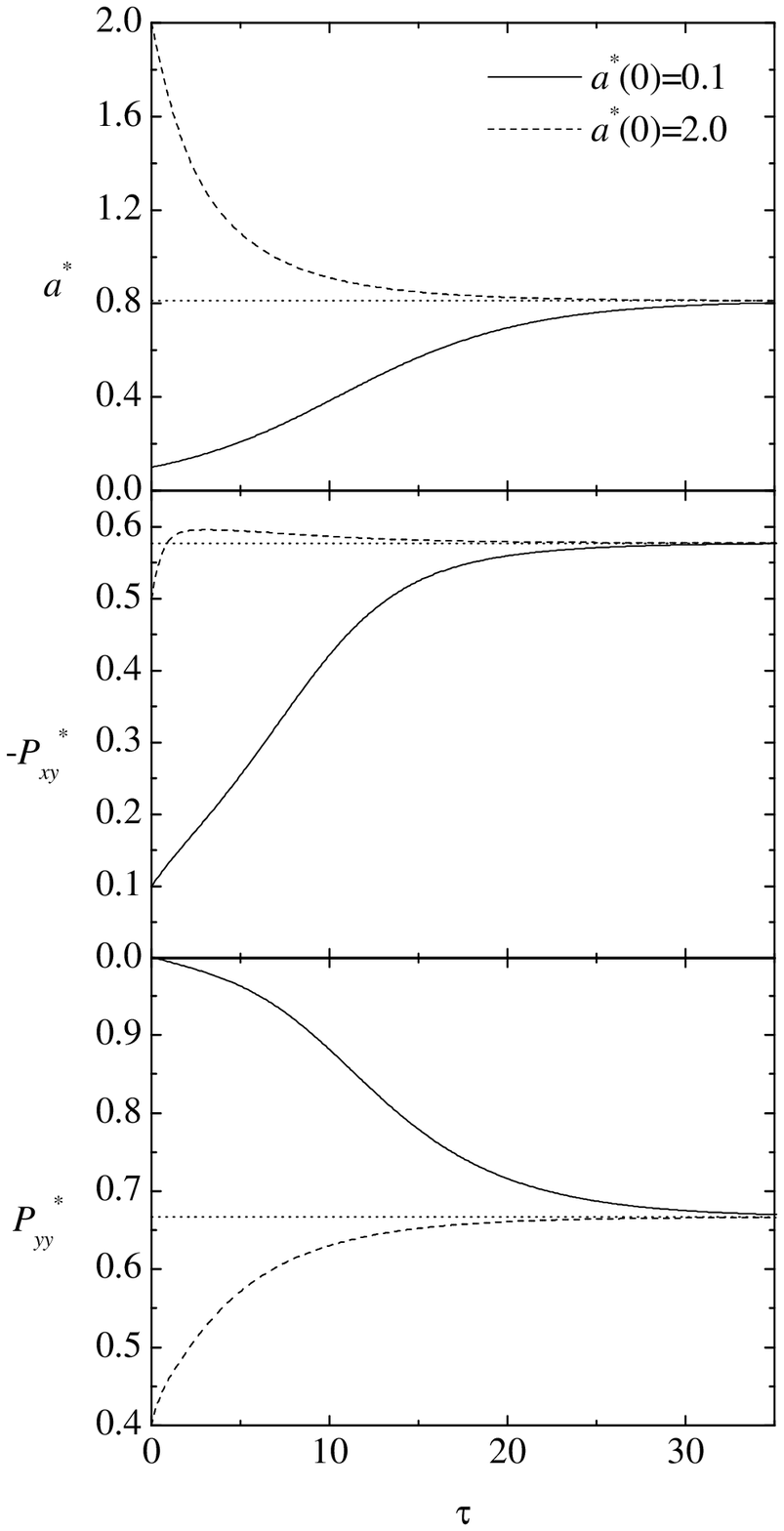}   
\caption{ Time evolution of the reduced shear rate $a^*(\tau)$, the reduced shear stress $-P_{xy}^*(\tau)$, and the    
reduced normal stress $P_{yy}^*(\tau)$ for $d=3$, $\alpha=0.5$, and two different initial conditions:  (i)    
$a^*(0)=0.1$, $P_{xy}^*(0)=-0.1$,  $P_{yy}^*(0)=1$ (solid lines) , and (ii) $a^*(0)=2$, $P_{xy}^*(0)=-0.5$,     
$P_{yy}^*(0)=0.4$ (dashed lines). The horizontal dotted lines represent the steady state values $a^*_s=0.812$,    
$-P_{xy,s}^*=0.577$, and $P_{yy,s}^*=0.667$.    
\label{fig4}}   
\end{figure}

In order to get the branch of the hydrodynamic solution corresponding to $   
a^{\ast }<a_{s}^{\ast }$ one must apply the physical boundary condition    
\begin{equation}   
\lim_{a^{\ast }\rightarrow 0}\eta ^{\ast }(a^{\ast },\alpha )=\eta   
_{0}^{\ast }\left( \alpha \right) ,\quad  \lim_{a^{\ast }\rightarrow   
0}P_{yy}^{\ast }(a^{\ast },\alpha )=1.  \label{3.12}   
\end{equation}   
In practice, the branch $a^*< a_s^*$ for each curve of Fig.\ \ref{fig2} has been obtained by   
solving   
numerically the coupled set (\ref{3.10}) and (\ref{3.11}) with the initial   
condition   
\beq   
P_{xy}^*=-\eta_0 a^*(0),\quad P_{yy}^*=1   
\label{n6}   
\eeq   
with $a^{\ast }(0)=10^{-5}$.

As said before, the branch corresponding to $a^*>a_s^*$ is inaccessible starting from an initial value $a^*(0)<a_s^*$.    
Therefore, in order to obtain $\eta^*(a^*)$ for $a^{\ast }>a_{s}^{\ast }$, one must take the  boundary   
condition at the point at infinity.    
{ For that case, simple physical intuition is not enough to determine the appropriate boundary conditions, so one    
must resort to a detailed asymptotic analysis  of Eqs.\ (\ref{3.10}) and (\ref{3.11}). Such an analysis is carried out    
in Appendix \ref{appB}, where we find that that $-P_{xy} ^{\ast }\sim a{^{\ast }}^{-1/3}$, $P_{yy}^{\ast   
}\sim a{^{\ast }}^{-2/3}$ for asymptotically large $a^{\ast }$.} More   
specifically, if the shear rate $a^{\ast }$ is much larger than   
1, then    
\begin{subequations}   
\beq   
P_{xy}^*=-\left( \frac{9d^2\beta}{56}\right) ^{1/3}{a^{\ast }}^{-1/3},   
\label{3.16a}   
\eeq   
\beq   
P_{yy}^{\ast }=   
\left( \frac{21d\beta^2}{64}\right) ^{1/3}{a^{\ast }}^{-2/3}.     
\label{3.16b}   
\eeq   
\label{3.16}   
\end{subequations}   
The   
 branch $a^*>a^*_s$ in Fig.\ \ref{fig2} has been obtained from the numerical solution of the coupled set (\ref{3.10})    
and (\ref{3.11}) with the initial   
conditions (\ref{3.16}) with $a^*(0)=10$.   
   
\section{Discussion\label{sec4}}   
   
The primary observation here has been that granular fluids admit   
hydrodynamic steady states that are inherently beyond the scope of the   
Navier--Stokes or Newtonian hydrodynamic equations. The reason for this is   
the existence of an internal mechanism, collisional cooling, that sets the   
scale of the spatial gradients in the steady state. For normal fluids,   
this scale is set by external sources (boundary conditions, driving forces)   
that can be controlled to admit the conditions required for Navier--Stokes   
hydrodynamics. In contrast, collisional cooling is fixed by the mechanical   
properties of the particles making up the fluid. 
An
example is a sheared fluid where the work done at the boundaries is
balanced by a combination of collisional cooling and  internal heat
flux. To illustrate and emphasize these effects in more detail and
quantitatively, we have considered the idealized state of uniform shear flow (USF), for which no
heat flux occurs. The Lees--Edwards boundary conditions generating USF cannot be applied in experimental conditions, although they can be easily implemented in computer simulations.
However, even for the laboratory conditions of Couette
flow the  cooling rate will still be a function of the hydrodynamic
gradients leading to similar non-Newtonian behavior.

{ In spite of the extensive prior work on USF   
for granular fluids, the observation about its inherent non-Newtonian character is apparently new although implicit in results obtained   
from various models (e.g., the observation that ${a^*}^2 \sim 1-\alpha$)}. This is significant   
because molecular dynamics simulation of steady USF has been used for real fluids to measure the Newtonian shear viscosity.   
The analysis here shows that this is not possible for granular fluids. More generally, it   
provides a caution regarding the simulation of other steady states to study Navier--Stokes   
hydrodynamics when the gradients are strongly correlated to the collisional cooling. On a more   
positive note, these results show that USF is an ideal testing ground for the study of rheology   
since any choice of the shear rate and $\alpha$ will provide non-Newtonian effects. It is one of the    
fascinating features of granular fluids that phenomena associated with complex fluids are   
more easily accessible than for simple atomic fluids \cite{G01}.}    
   
The macroscopic state of   
steady uniform shear flow is simple in the sense that it is   
spatially uniform in the local Lagrangian frame, and the only hydrodynamic   
gradient is that of the velocity field characterized by the scalar shear   
rate $a$. Non-Newtonian effects occur through the dependence of the shear   
viscosity on the shear rate. The condition for a steady state is an exact   
balance between viscous heating and collisional cooling. It is shown in   
Section \ref{sec2} that this implies a value of the appropriate dimensionless shear   
rate that is fixed by the restitution coefficient. Hence, it is not possible   
to make this hydrodynamic gradient small, as required for the Newtonian   
limit, by any initial control of $a$ or the initial temperature. Further   
quantitative illustration of this is given in Section \ref{sec3} where the shear rate   
dependence of the shear viscosity is determined from { an approximate Grad's solution of the Boltzmann equation.} The    
{ temporal} approach to the steady state via cooling or heating   
proceeds along a rheological equation of state shown in Fig.\ \ref{fig2}, but the final steady state is independent of    
the initial conditions and   
lies in the non-Newtonian domain for any value of the restitution   
coefficient $\alpha \neq 1$. One consequence is that experimental or   
simulation measurements in steady USF provide no information about   
Navier--Stokes transport.

This phenomenon of peculiar steady states for granular fluids extends to   
other states as well. The steady state macroscopic balance equations are   
\begin{equation}   
\nabla \cdot \left( n\mathbf{u}\right) =0,  \label{4.1}   
\end{equation}   
\begin{equation}   
-\mathbf{u}\cdot \nabla \ln T-\frac{2}{dnT}\left( P_{ij}\nabla   
_{j}u_{i}+\nabla \cdot \mathbf{q}\right) =\zeta ,  \label{4.2}   
\end{equation}   
\begin{equation}   
\mathbf{u}\cdot \nabla u_{i}+(mn)^{-1}\nabla _{j}P_{ij}=0.  \label{4.3}   
\end{equation}   
In the absence of collisional cooling the spatial gradients can be sustained   
only by boundary conditions or external forces. However, for finite $\zeta $   
spatial gradients can exist that are not controlled by such external   
sources and their size may preclude the validity of simple Navier--Stokes   
hydrodynamics. Each physical state should be checked for this possibility.   
   
{ As another simple example, consider the steady state of a granular gas enclosed between two parallel plates at rest    
and maintained at the same temperature. For elastic particles such a steady state is trivially that of equilibrium at    
the wall temperature. Nevertheless, for a granular gas the collisional dissipation induces a heat flux, cf.\ Eq.\    
(\ref{4.2}), and hence a thermal gradient, along with a density gradient, exists \cite{BC98,SMR99,RRSC00}. In the    
quasielastic limit, the kinetic model of the Boltzmann equation proposed in Ref.\ \cite{brey99model3} admits a    
solution characterized by a constant pressure $p=nT$ and a heat flux given by \cite{GSD}   
\beq   
\textbf{q}=-\frac{d+2}{2}\frac{p}{m\nu}\lambda_s^*(\alpha)\nabla T,   
\label{4.4}   
\eeq   
where   
\beq   
\lambda_s^*(\alpha)\to 1+\frac{29d^2+116d-28}{4d(d+2)}(1-\alpha).   
\label{4.5}   
\eeq   
On the other hand, the expression for the heat flux in the Navier--Stokes order given by the Chapman--Enskog expansion    
is \cite{BDKyS98,GD99,BC01}   
\beqa   
\textbf{q}&=&-\lambda_0(\alpha)\nabla T-\mu_0(\alpha) \nabla n\nn   
&=&-\frac{d+2}{2}\frac{p}{m\nu}\bar{\lambda}_0^*(\alpha)\nabla T,   
\label{4.6}   
\eeqa   
where $\lambda_0$ is the  thermal conductivity, $\mu_0$ is a transport coefficient with no analog in the elastic case,    
and in the last step we have taken into account that $\nabla p=0$, so that $\nabla n=-(n/T)\nabla T$. The expression    
for the dimensionless effective thermal conductivity $\bar{\lambda}_0^*(\alpha)$ in the quasielastic limit is   
\beq   
\bar{\lambda}_0^*(\alpha)\to 1+\frac{3d-4}{4d}(1-\alpha).   
\label{4.7}   
\eeq   
Equations (\ref{4.5}) and (\ref{4.7}) clearly show that ${\lambda}_s^*(\alpha)\neq \bar{\lambda}_0^*(\alpha)$. In the    
three dimensional case, for instance, the coefficient of $1-\alpha$ in ${\lambda}_s^*(\alpha)$ is about twenty times    
larger than that of $\bar{\lambda}_0^*(\alpha)$. Therefore, the Navier--Stokes transport coefficients do not describe    
correctly the heat flux in the steady state, except again in the trivial case of elastic collisions. }   
   
\begin{acknowledgments}   
A.S. and V.G. acknowledge partial support from the Ministerio de   
Ciencia y Tecnolog\'{\i}a   
 (Spain) through grant No.\ FIS2004-01399.   
The research of J.W.D. was supported by Department of Energy Grants   
DE-FG03-98DP00218 and DE-FG02ER54677.   
\end{acknowledgments}

\appendix   
\section{Linear stability analysis of the steady state solution\label{appA}}   
In this Appendix we carry out a linear stability analysis of the steady state solution (\ref{3.7a})--(\ref{3.8a}) of    
the set  of evolution equations (\ref{3.3})--(\ref{3.5}). First, we write   
\beqa   
a^*(\tau)&=& a^*_s+\delta a^*(\tau),\quad P_{xy}^*(\tau)=P_{xy,s}^*+\delta P_{xy}^*(\tau),\nn   
 P_{yy}^*(\tau)&=&P_{yy,s}^*+\delta P_{yy}^*(\tau).   
\label{A.1}   
\eeqa   
Substituting (\ref{A.1}) into Eqs.\ (\ref{3.3})--(\ref{3.5}) and neglecting nonlinear terms, one gets   
\beq   
\partial_\tau \left(   
\begin{array}{c}   
\delta a^*\\   
\delta P_{xy}^*\\   
a^*_s\delta P_{yy}^*   
\end{array}   
\right)=-\mathsf{L}\cdot \left(   
\begin{array}{c}   
\delta a^*\\   
\delta P_{xy}^*\\   
a^*_s\delta P_{yy}^*   
\end{array}   
\right),   
\label{A.2}   
\eeq   
where $\mathsf{L}$ is the matrix   
\beq   
\mathsf{L}=\left(   
\begin{array}{ccc}   
{\zeta_0^*}/{2}&-   
{\zeta_0^*(\beta+\zeta_0^*)^2}/{2\beta}&0\\   
{\beta^2}/{(\beta+\zeta_0^*)^2}&\beta+2\zeta_0^*&1\\   
{\beta \zeta_0^*}/{(\beta+\zeta_0^*)}&-\zeta_0^*(\beta+\zeta_0^*)&\beta+\zeta_0^*   
\end{array}   
\right).   
\label{A.3}   
\eeq   
In Eq.\ (\ref{A.3}) use has been made of the explicit expressions (\ref{3.7a})--(\ref{3.8a}) for the steady state    
solution.   
The time evolution of the deviations from the steady solution is governed by the eigenvalues of $\mathsf{L}$. If the    
real parts of those eigenvalues are positive the steady solution is linearly stable, while it is unstable otherwise.   
The solution of the characteristic equation $\det(L_{ij}-\ell\delta_{ij})=0$ yields a real eigenvalue $\ell_1$ and a    
pair of complex conjugate eigenvalues $\ell_2$, $\ell_3$. We have verified that $\ell_1$ and the real parts of    
$\ell_{2,3}$ are positive definite for all values of the coefficient of restitution $\alpha<1$, $\ell_1$ being smaller    
than $\text{Re}~ \ell_{2,3}$. Consequently, the steady state solution is stable, and the characteristic relaxation    
time (measured by the number of collisions) is $\ell_1^{-1}$.   
As an illustration, Fig.\ \ref{fig3} shows the $\alpha$-dependence of $\ell_1$ and $\text{Re}~\ell_{2,3}$ for $d=3$.    
It is interesting to note that $\ell_1\to 0$ in the elastic limit $\alpha\to 1$. This is a consequence of the fact    
that there is no steady solution at $\alpha=1$, namely $a^*(\tau)$ decays algebraically as $a^*(\tau)\approx    
(2\tau/d)^{-1/2}$.   
\begin{figure}[tbp]   
\includegraphics[width=.90 \columnwidth]{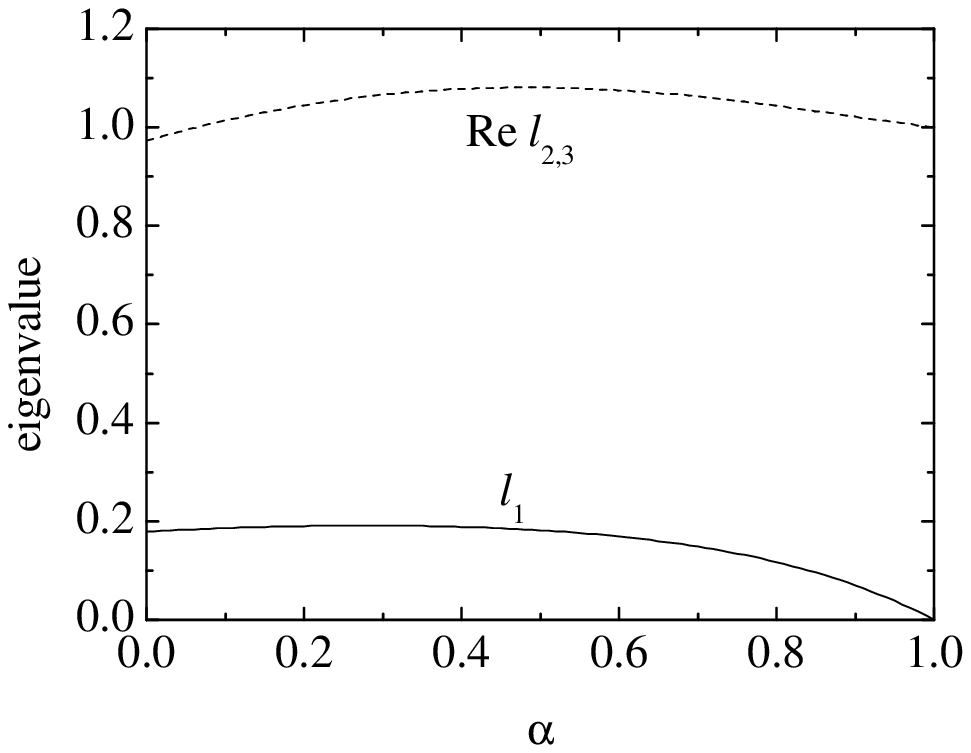}   
\caption{ Plot of the eigenvalue $\ell_1$ and of the real part of $\ell_{2,3}$ as a function of the coefficient of    
restitution in the three-dimensional case.   
\label{fig3}}   
\end{figure}   
   
The above stability analysis is restricted to small deviations from the steady state solution. The proof that the    
steady state solution is  an attractor as $\tau\to \infty$ for any uniform (in the Lagrangian frame) initial condition    
seems to be quite difficult, so that one has to resort to numerical evidence. As was illustrated in Fig.\ \ref{fig4},    
the numerical results confirm that the steady state is indeed an attractor.

\section{Hydrodynamic solution for large shear rates\label{appB}}   
In this Appendix we will get the asymptotic behavior of the solution to Eqs.\ 	(\ref{3.10}) and (\ref{3.11}) for    
large values of the reduced shear rate $a^*$. On physical grounds, the asymptotic solution must have the general form   
\beq   
P_{xy}^*\approx -A_1 {a^*}^{-b_1}, \quad P_{yy}^*\approx A_2 {a^*}^{-b_2},   
\label{B.1}   
\eeq   
where $A_1>0$, $A_2>0$, and  $b_2\geq 0$. Inserting (\ref{B.1}) into Eqs.\ (\ref{3.10}) and (\ref{3.11}), one has   
\beq   
\left[2\beta-b_1\zeta_0^*+\frac{2(b_1+2)}{d}A_1 {a^*}^{1-b_1}\right]A_1 {a^*}^{-1-b_1}=2A_2 {a^*}^{-b_2},   
\label{B.2}   
\eeq   
\beq   
\left[2\beta-b_2\zeta_0^*+\frac{2(b_2+2)}{d}A_1 {a^*}^{1-b_1}\right]A_2 {a^*}^{-b_2}=2\beta .  
\label{B.3}   
\eeq   
In order to carry out a balance analysis of the leading terms in (\ref{B.2}) and (\ref{B.3}), it is convenient to    
consider separately the cases $b_1>1$, $b_1=1$, and $b_1<1$. If $b_1>1$, then ${a^*}^{1-b_1}\to 0$ and Eqs.\    
(\ref{B.2}) and (\ref{B.3}) become   
\beq   
\left(2\beta-b_1 \zeta_0^*\right) A_1{a^*}^{-1-b_1}=2A_2{a^*}^{-b_2},   
\label{B.4}   
\eeq   
\beq   
\left(2\beta-b_2 \zeta_0^*\right) A_2{a^*}^{-b_2}=2\beta.   
\label{B.5}   
\eeq   
Equation (\ref{B.4}) implies that $b_2=b_1+1>2$, while Eq.\  (\ref{B.5}) yields $b_2=0$ and $A_2=1$. Therefore, Eqs.\    
(\ref{B.4}) and (\ref{B.5}) are mutually inconsistent.   
The next possibility is $b_1=1$. This gives   
\beq   
\left(2\beta-\zeta_0^*+\frac{6}{d}A_1\right)A_1 {a^*}^{-2}=2A_2 {a^*}^{-b_2},   
\label{B.6}   
\eeq   
\beq   
\left[2\beta-b_2\zeta_0^*+\frac{2(b_2+2)}{d}A_1\right]A_2 {a^*}^{-b_2}=2\beta.   
\label{B.7}   
\eeq   
According to Eqs.\ (\ref{B.6}) and (\ref{B.7}), $b_2=2$ and $b_2=0$, respectively, what is again inconsistent.   
   
Finally, we consider the case $b_1<1$, so that Eqs.\ (\ref{B.2}) and (\ref{B.3}) reduce to   
\beq   
\frac{b_1+2}{d}A_1^2{a^*}^{-2b_1}=A_2{a^*}^{-b_2},   
\label{B.8}   
\eeq   
\beq   
\frac{b_2+2}{d}A_1A_2{a^*}^{1-b_1-b_2}=\beta.   
\label{B.9}   
\eeq   
These two equations are consistent provided that   
\beq   
b_1=\frac{1}{3},\quad b_2=\frac{2}{3},   
\label{B.10}   
\eeq   
\beq   
A_1=\left( \frac{9d^2\beta}{56}\right) ^{1/3},\quad    
A_2=\left( \frac{21d\beta^2}{64}\right) ^{1/3}.    
\label{B.11}   
\eeq    
   
Therefore, for large reduced shear rates,   
\beq   
P_{xy}^*\approx-\left( \frac{9d^2\beta}{56}\right) ^{1/3}{a^{\ast }}^{-1/3},   
\label{B.12}   
\eeq   
\beq   
P_{yy}^{\ast }\approx   
\left( \frac{21d\beta^2}{64}\right) ^{1/3}{a^{\ast }}^{-2/3}.     
\label{B.13}   
\eeq   
It is interesting to remark that $|P_{xy}^*|\to 0$ both for small and large shear rates. Thus, $|P_{xy}^*|$ exhibits a    
non-monotonic dependence on $a^*$ characterized by a maximum value $|P_{xy}^*|_{\text{max}}$ at a certain value    
$a^*_{\text{max}}$ of the reduced shear rate. For instance, in the three-dimensional case we have found    
$(a^*_{\text{max}},|P_{xy}^*|_{\text{max}})=(0.64,0.634)$, $(1.22,0.595)$, and $(1.81,0.579)$ for $\alpha=0$, $0.5$,    
and $0.9$, respectively. A similar non-monotonic behavior of  $|P_{xy}^*|$ occurs in the elastic case as well    
\cite{GS03}.

\end{document}